%Paper: hep-ph/9410203
%From: seiberg@physics.rutgers.edu (Nathan Seiberg)
%Date: Mon, 3 Oct 94 14:19:46 EDT

\input harvmac

\def\np#1#2#3{Nucl. Phys. {\bf B#1} (#2) #3}
\def\pl#1#2#3{Phys. Lett. {\bf #1B} (#2) #3}
\def\prl#1#2#3{Phys. Rev. Lett. {\bf #1} (#2) #3}
\def\physrev#1#2#3{Phys. Rev. {\bf D#1} (#2) #3}

\def\Tr{{\rm Tr ~}}

\Title{hep-ph/9410203, RU-94-75, IASSNS-HEP-94/79}
{\vbox{\centerline{Proposal for a Simple Model of}
\centerline{Dynamical SUSY Breaking}}}
\bigskip
\centerline{K. Intriligator$^1$, N. Seiberg$^{1,2}$, and S. H.
Shenker$^1$}
\vglue .5cm
\centerline{$^1$Department of Physics and Astronomy}
\centerline{Rutgers University}
\centerline{Piscataway, NJ 08855-0849, USA}
\vglue .3cm
\centerline{$^2$ School of Natural Sciences}
\centerline{Institute for Advanced Study}
\centerline{Princeton, NJ 08540, USA}

\medskip

\noindent
We discuss supersymmetric $SU(2)$ gauge theory with a single matter field
in the $I=3/2$ representation.  This theory has a moduli space of
exactly degenerate vacua.  Classically it is the complex plane with an
orbifold singularity at the origin.  There seem to be two possible
candidates for the quantum theory at the origin.  In both the global
chiral symmetry is unbroken. The first is interacting quarks and gluons
at a non-trivial infrared fixed point -- a non-Abelian Coulomb phase.
The second, which we consider more likely, is a confining phase where
the singularity is simply smoothed out.  If this second, more likely,
possibility is realized, supersymmetry will dynamically break when a
tree level superpotential is added.  This would be the simplest known
gauge theory which dynamically breaks supersymmetry.

\Date{10/94}
%\draftmode

Only supersymmetric gauge theories with chiral matter content can
dynamically break supersymmetry with a stable vacuum.  The reason for
this is that the matter fields in a non-chiral gauge theory can be
given mass terms and decoupled, leaving as the low energy limit a
supersymmetric pure gauge theory which is known to have non-zero
Witten index ($\Tr (-1)^F\neq 0$) and, hence, unbroken supersymmetry
\ref\windex{E. Witten, \np{202}{1982}{253}.}.
As long as the theory has a stable vacuum, $\Tr(-1)^F$ is independent
of the magnitude of the added mass terms and so the theory with
non-chiral matter does not break supersymmetry \windex.  Chiral models
with dynamical supersymmetry breaking were given in
\ref\ads{I. Affleck, M. Dine, and N. Seiberg, \pl{137}{1984}{187};
\prl{52}{1984}{493}; \pl{140}{1984}{59};
\np{256}{1985}{557}.}.
Here we will discuss a simpler model, based on an $SU(2)$ gauge theory
with chiral matter content, which may well break supersymmetry upon
adding a tree level superpotential.

Consider supersymmetric $SU(2)$ gauge theory with a single matter
field $Q$ in the $I=3/2$ representation.  The quadratic index $\mu$ of
the $I=3/2$ representation of $SU(2)$ is $\mu = 10$ (for general $I$
it is $\mu =2(2I+1)I(I+1)/3$).  Since this index is even, the theory
is well-defined; it does not suffer {}from the global anomaly of
\ref\wganom{E. Witten, \pl{117}{1982}{324}.}.
In addition, this theory is asymptotically free because the matter
content satisfies the relevant inequality: $\mu <12$.  Therefore, this
is a sensible theory containing a single pseudo-real field.  By Bose
statistics in the superfield $Q$, there is no non-zero gauge singlet
which is quadratic in $Q$ -- it is impossible to give the pseudo-real
field $Q$ a mass.  In other words, this theory is chiral.

The basic gauge singlet superfield is $u=Q^4$, with a totally
symmetric contraction of the gauge indices.  At the classical level,
this theory has a moduli space of degenerate vacua labeled by the
expectation value of $u$.  For $u\neq 0$ the $SU(2)$ gauge group is
completely broken by the Higgs mechanism.  The classical Kahler
potential for $u$ is $K_{cl}=QQ^\dagger \sim (uu^\dagger)^{1/4}$,
which is singular at $u=0$.  This is a $Z_4$ orbifold singularity.  It
reflects the fact that the gauge bosons become massless at $u=0$ and
hence must be included in the low energy effective theory.

To analyze the theory at the quantum level, note that there is an
anomaly free $U(1)_R$ symmetry under which the scalar component of $Q$
has charge $q=3/5$.  This symmetry, along with holomorphy, restricts
the form of an effective superpotential to be $W_{eff}=au^{5/6}\Lambda
^{-1/3}$, where $\Lambda$ is the dynamically generated scale of the
theory and $a$ is some constant.  However, this superpotential does
not have sensible behavior in the weak coupling $u\gg \Lambda$ limit
and, hence, $a=0$.  Therefore, the quantum theory also has a moduli
space of degenerate vacua.

The most interesting question about the quantum moduli space is the
nature of the classical singularity at $u=0$ in the quantum theory.
Similar spaces in other theories have been analyzed and several
possibilities found.  First, a classical singularity can be smoothed
out in the quantum theory
\nref\nati{N. Seiberg, hep-th/940244, \physrev{49}{1994}{6857}.}%
\nref\sw{N. Seiberg and E. Witten, hep-th/9407087, \np{426}{1994}{19};
hep-th/9408099, RU-94-60, IASSNS-HEP-94/55.}%
\refs{\nati, \sw}.  Alternatively, the singularities in the quantum
theory are associated with new massless fields which are collective
excitations of the elementary fields \refs{\nati, \sw}.  Finally, the
quantum theory at the singularity can be similar to the classical
theory with massless interacting quarks and gluons \nati\ -- a
non-Abelian Coulomb phase.  (Of course, other as yet unknown
possibilities might also exist.)

In the example at hand there seem to be two plausible alternatives for
the quantum behavior at $u=0$.  First, there could be a non-Abelian
Coulomb phase.  The beta function is
\eqn\betafun{\beta(\alpha)= - {\alpha^2 \over 2 \pi} \left(1- {71
\over 4 \pi}\alpha + \CO(\alpha^2)\right)}
where $\alpha = g^2/4\pi$.  Ignoring the higher order corrections,
there is a non-trivial infrared stable fixed point at $\alpha^{*} =
4\pi/71$.  The appropriate effective coupling might well be larger
because of the large matter Casimir.  This coupling may be small
enough for perturbation theory to be reliable and the nontrivial fixed
point and its non-Abelian Coulomb behavior to be physical.  We have no
compelling argument that this is indeed the case.

There is second more interesting dynamical possibility for the
behavior at $u=0$ which we find more likely.  The topology of
the space of vacua could remain the entire $u$ plane and the classical
singularity in the Kahler potential (metric) at the origin could be
smoothed out.  The massless spectrum on the whole moduli space would
consist just of the $u$ quanta.  This possibility passes the following
highly non-trivial consistency test.  At the point $u=0$ the global
$U(1)_R$ symmetry is unbroken.  The massless spectrum -- the $u$
quanta -- should saturate the 't Hooft anomaly conditions
\ref\thooft{G. 't Hooft, in {\it Recent Developments in Gauge Theories},
eds. G. 't Hooft {\it et al.} (Plenum Press, New York, 1980)}.
The fermions in the microscopic theory are the 3 gluinos, each with
$R$ charge 1, and the 4 quark components of $Q$, each with $R$ charge
$-2/5$.  In the conjectured macroscopic theory there is a single
fermion, the fermionic component of $u$, which has $R$ charge $7/5$.
There are two equations to check:
\eqn\anommatch{\eqalign{&\Tr R = 3+4(-2/5)=7/5\cr
&\Tr R^3= 3+4(-2/5)^3=(7/5)^3,}}
which are indeed satisfied.  This seems too miraculous to be a
coincidence.  Therefore in
the remainder of this paper we will assume that the theory
behaves in this manner and proceed to explore the consequences.

By the R symmetry the Kahler potential is of the form $K=
|\Lambda|^{2} k(uu^\dagger/ |\Lambda |^{8})$ for some function $k$.
The smooth behavior near the origin and the semiclassical behavior at
infinity constrain $K$ to satisfy
\eqn\kbehavior{K=|\Lambda |^{2}
k(uu^\dagger/ |\Lambda |^{8}) \sim
\cases{ uu^\dagger |\Lambda |^{-6}& for $uu^\dagger \ll  \Lambda^8 $ \cr
(uu^\dagger)^{1/4} & for  $uu^\dagger \gg \Lambda^8. $ \cr}}
Since the low energy superpotential vanishes and the Kahler potential
is independent of the phase of $u$, the low energy theory has, in
addition to the $U(1)_R$ symmetry, an accidental global $U(1)$
symmetry under which the chiral superfield $u$ has, say, charge $1$.
The corresponding symmetry in the microscopic theory is anomalous.  In
the macroscopic theory it is violated by higher dimensional terms
which are not determined by the Kahler potential and the
superpotential.

The fact that the massless spectrum at $u=0$ consists only of $u$
means that there is confinement there even though the chiral $U(1)_R$
symmetry is unbroken.  Confinement without chiral symmetry breaking is
in contrast to the prediction of the most attractive channel (MAC)
arguments.  These arguments would have suggested that a bilinear
condensate is formed, breaking the gauge symmetry $SU(2)\rightarrow
U(1)$ with chiral symmetry breaking at a scale much larger that
$\Lambda$.  However, we see no sign of a massless photon associated
with an unbroken $U(1)$.  Other counterexamples to the MAC intuition
have also been observed in other supersymmetric gauge theories \nati .

We now consider perturbing the theory by a tree level superpotential
\eqn\treelevs{W_{tree}=\lambda u.}
This term is non-renormalizable; perhaps it can be realized as the low
energy effective superpotential obtained {}from a larger renormalizable
gauge theory at a scale $m \sim \lambda^{-1}$, or as a low energy
limit of string theory with $\lambda \sim 1/M_p$.  In any event, our
effective theory (as well as the microscopic gauge theory) is then
only sensible for $|u|<|\lambda|^{-4}$.  For $|u|>|\lambda|^{-4}$ we
would have to include the fields in a more fundamental renormalizable
theory and the physics of this region would contain non-universal
dependence on the choice of the higher energy theory.  We can take
$\lambda \ll \Lambda ^{-1}$ so as to have room to take $u$ outside the
region of strong coupling gauge dynamics, $|u|\gg |\Lambda |^4$,
without running into the scale of the non-renormalizable term.

Using symmetries and holomorphy  as in
\ref\nonren{N. Seiberg, hep-ph/9309335, \pl{318}{1993}{469}.},
the exact superpotential of the low energy theory is constrained to be
\eqn\wsymm{W_{exact}=\lambda u f(t=\lambda ^6\Lambda ^2 u).}
Perturbative behavior in $\lambda$ and the requirement of no cuts in
$u$ leads to $f=\sum _{n=0}^\infty a_nt^n$; the $n$-th term has the
quantum numbers of a $2n$ instanton contribution.  In our allowed
range of $u$, $|t|\ll 1$ and so we have $W\approx \lambda u$.
Therefore, the scalar potential in our region is
\eqn\vis{V_{eff}=(K_{uu^\dagger})^{-1} |W_u|^2=
(K_{uu^\dagger})^{-1}|\lambda|^2.}
Using the asymptotic behavior \kbehavior, it is clear that
supersymmetry is broken with vacuum energy
\eqn\vacen{E  \sim |\Lambda |^6|\lambda |^2.}

Note that the superpotential $W=\lambda u$ preserves an $R$ symmetry
which is a combination of the anomaly free $U(1)_R$ symmetry and the
accidental $U(1)$ symmetry.  Therefore, the supersymmetry breaking
found here is in accord with the analysis of
\ref\nands{A. Nelson and N. Seiberg, hep-ph/9309299,
\np{416}{1994}{46}}.
If the minimum energy vacuum is at $u=0$, this $R$ symmetry is
unbroken.  Otherwise, there is a massless Goldstone boson.  If we
include the higher terms in \wsymm , the theory will have a
supersymmetric minimum, but outside the domain $|u|<|\lambda |^{-4}$
where this description is valid.  Again, the existence of such a
minimum is in accord with \nands\ because with the higher terms in
\wsymm\ included there will no longer be an $R$ symmetry.

To summarize, classically the superpotential $\lambda u$ does not lead to
spontaneous supersymmetry breaking because of the singularity of the
Kahler potential at $u=0$.  Quantum mechanically the singularity is
smoothed out and supersymmetry is dynamically broken.  We would like to
conclude by stressing that this mechanism for supersymmetry breaking is
more general than the specific model studied here. It can no doubt exist
in other models and perhaps even in Nature.

\centerline{{\bf Acknowledgments}}

We would like to thank M.R. Plesser and E. Witten for useful discussions.
This work was supported in part by DOE grant \#DE-FG05-90ER40559.

\listrefs

\end